\documentclass[12pt]{article}
\usepackage{amsfonts,amsmath,amssymb,epsf,cite}
\usepackage{graphicx,color}
\usepackage[usenames,dvipsnames]{pstricks}
 \def\m{{\mu}}

 \def\frac#1#2{{#1\over #2}}
 
\def\tch{\tilde{\chi}}

 \def\ch{{\chi}}

\def\cN{{\cal N}}

\def\cD{{\cal D}}
\def\ttxi{\tilde{\xi}}

 \def\CN{{\cal N}}
\def\cL{{\cal L}}
\def\cP{{\cal P}}
\def\cO{{\cal O}}

 \def\ti{\tilde}

 \def\no{\nonumber \\}
\def\nn{\nonumber}

 \def\ov{\overline}
 
 \def\vp{\varphi}

 \def\m{{\mu}}

 \def\frac#1#2{{#1\over #2}}

 \def\ch{{\chi}}

 \def\CN{{\cal N}}

\def\be{\begin{equation}}
\def\ee{\end{equation}}
\def\ba{\begin{eqnarray}}
\def\ea{\end{eqnarray}}
\def\vx{{\vec{x}}}

\def\ts{\tilde{s}}
\def\ben{\begin{equation}}
\def\een{\end{equation}}
\def\bea{\begin{eqnarray}}
\def\eea{\end{eqnarray}}

\begin{document}
\begin{titlepage}
\thispagestyle{empty}
\begin{flushright}
UK/11-08
\end{flushright}

\bigskip

\begin{center}
\noindent{\Large \textbf
{Holographic Quantum Quench \footnote{Based on talks at 11th
    Workshop on Non-perturbative QCD (Paris, June 2011), Sixth Crete
    Regional Meeting on String Theory (Milos, June 2011) and Quantum
    Theory and Symmetries 7 (Prague, August 2011)}} }\\
\vspace{2cm} \noindent{
Sumit R. Das\footnote{e-mail:das@pa.uky.edu}}\\
\vspace{1cm}
Department of Physics and Astronomy, \\
 University of Kentucky, Lexington, KY 40506, USA
\end{center}

\vspace{0.3cm}
\begin{abstract}

We discuss recent results in the study of the evolution of strongly
coupled field theories in the presence of time dependent couplings
using the holographic correspondence.  The aim is to understand (i)
thermalization and (ii) universal behavior when the coupling crosses a
critical point. Our emphasis is on situations where a subset of bulk
fields can be treated in a probe approximation. We consider two
different setups. In the first, defect conformal field theories are
described by probe branes in AdS space-times, and an initial vacuum
state evolves due to a time dependent coupling in the probe
sector. While a black hole formation is invisible in this
approximation, we show that thermalization can nevertheless happen -
this is signalled by formation of apparent horizons on the brane
worldvolume. In the second setup, we consider a probe bulk scalar
field in the background of a AdS black brane. In equilibrium, this
system undergoes a critical phase transition at some temperature when
the source for the dual operator vanishes. For a time dependent source
which goes across this critical point, we show that a zero mode of the
bulk field dominates the dynamics and leads to scaling behavior of the
order parameter as a function of the rate of change.

\end{abstract}
\end{titlepage}

\newpage

\section{Introduction}
\vspace{0.5cm}
The problem of quantum quench, i.e. the response of a system to a time
dependent coupling, 
has recently attracted a lot of attention in several areas of
many-body physics, particularly because of progress in cold atom
experiments \cite{sengupta},\cite{CCa},\cite{CCc}. This problem is
interesting for at least two reasons. The first relates to the
question of thermalization. Suppose we start with the ground state and
then turn on a time dependent coupling which again approaches a
constant at late times. Does the system evolve into some kind of
steady state ? If so, is the state "thermal" in any sense ?  

The second question deals with the situation where the quench takes
place across a value of the parameter where there is an equilibriium
critical point. In this case there is some evidence that the time
evolution carries some universal features of the critical point.
Assuming that there is only one scale which governs the behavior of
the system in the critical region, one can, e.g. derive scaling
properties of one point functions following adaptations of the early
work of Kibble and Zurek \cite{kibble,zurek}. Suppose the coupling
approaches the critical coupling linearly, i.e.  
\ben 
(g-g_c) \sim vt
\een 
These arguments then show that the one point function of an
operator with conformal dimension $x$ at the critical point has a
scaling behavior \cite{sengupta}, \ben <\cO (t)> \sim
(v)^{\frac{x\nu}{z\nu+1}} F(tv^{\frac{z\nu}{z\nu+1}}) \een where $z$
is the dynamical critical exponent and $\nu$ is the correlation length
exponent. Another manifestation of such universal behavior appears in
$1+1$ dimensional field theories which are quenched {\em suddenly} to
a critical point. In this case, powerful methods of boundary conformal
field theory can be used to obtain the time dependence of correlation
functions \cite{CCa,CCc}. For example the one point function of a
generic operator with conformal dimension $x$ behaves as \ben <\cO(t)>
\sim e^{-\frac{\pi x t}{2\tau_0}} \een where $\tau_0$ denotes a length
scale which characterizes the initial theory away from
criticality. $\tau_0$ is of course not universal, so neither is the
relaxation time $\tau = 2\tau_0/(\pi x)$. However, the ratio of the
relaxation times for two {\em different} operators $\cO_1$ and $\cO_2$
with conformal dimensions $x_1$ and $x_2$ is universal, \ben
\frac{\tau_1}{\tau_2} = \frac{x_2}{x_1} \een Unlike equilibrium
critical phenomena there is no general theoretical framework to
understand such scaling relations.  In particular, there are very few
theoretical tools available to study such systems when they are
strongly coupled. It is, therefore, natural to explore if the AdS/CFT
correspondence \cite{Maldacena} - \cite{AdSR} is useful in this
problem.  

In this contribution we will summarize results obtained in this
approach. The work on thermalization is in collaboration with Tatsuma
Nishioka and Tadashi Takayanagi \cite{dnt}. The work on quench across
critical points is in in collaboration with Pallab Basu \cite{bdas}.

In the AdS/CFT correspondence, couplings of the boundary field theory
are boundary values of a bulk field. In the regime where supergravity
is valid, the problem of quantum quench then reduces to a classical
problem with given initial and boundary conditions. This problem has
been studied when the time dependent coupling is the boundary metric
or the gauge theory coupling (i.e. the boundary value of the
dilaton). Suppose this coupling is a constant in the far past and
future, and has a smooth time dependent profile at intermediate
times. In the bulk description this corresponds to a disturbance
created on the boundary which propagates in the bulk.  Under suitable
conditions, this leads to black hole formation in the bulk
\cite{janik,otherthermalization,holoentanglement}.  The correlators at
future time would then be thermal with a temperature characterized by
the Hawking temperature. The time scale after which this happens
depends on the nature of the correlators, but turns out to be always
smaller than what one would expect from a conformally invariant system
evolving to a thermal state. Thus, in this case thermalization of the
field theory is signalled by black hole formation. For homogeneous
planar collapse (i.e. a space independent coupling in Poincare patch),
a black hole is always formed. In the case of homogeneous collapse in
global AdS, a black hole is formed when the rate of change is fast
enough compared to the scale set by the radius of the sphere on which
the boundary theory lives.

In other situations, e.g. a slow variation of the coupling for global
AdS, a black hole is not formed so long as the supergravity
approximation is valid. Rather, if the coupling becomes {\em weak} at
some time, the bulk string frame curvature grows large, leading to a
breakdown of the supergravity approximation - thus mimicking a
space-like singularity \cite{Awad}. For the case of a slow variation
of the coupling it turns out that the gauge theory remains well
defined and may be used to show that a smooth passage through this
region of small coupling is possible without formation of a large
black hole.  Related scenarios appear in \cite{hertog} and \cite{eva}.

One of our main aims is to study quantum quench across critical
points. Many such critical points can be studied in setups where a
subset of the bulk fields can be treated in a probe approximation. In
this approximation, these probe fields provide the essential physics
and their backreaction to the background gravity can be ignored,
typically suppressed by $1/N$. This motivates us to study the problem
of quantum quench in situations where such a probe approximation is
valid.

We will first consider defect field theories which arise as dual
descriptions of a set of probe branes in the $AdS \times S$ bulk
\cite{KR}. This approach has been used extensively to study flavor
physics, as well as models with possible applications to condensed
matter systems. The nice feature of this approach is that the boundary
field theory is known, though they typically have supersymmetry. In
this case a quantum quench of couplings in this subsector becomes a
classical motion of these probe branes, with specified time dependent
boundary conditions at the $AdS$ boundary \cite{dnt}.   
We investigate the
question of thermalization in this context. Since the background
geometry is unchanged in this approximation any black hole which is
formed due to the quench is not visible. We will find that
thermalization is nevetheless visible - this manifests itself as the
formation of an apparent horizon on the brane worldvolume.

In the second setup we consider a "bottom-up" bulk theory of gravity
with a neutral scalar field \cite{liu1}, where one writes down a bulk
theory and {\em assume} that there is some dual field theory on the
boundary.  In this specific instance, 
the background is a $AdS_4$ charged black
brane and the coupling of the scalar is large, so that its
backreaction to the geometry is small. When the mass of the scalar
lies in the range $ -\frac{9}{4} <  m^2 < -\frac{3}{2}$ there is always
a critical temperature below which the trivial solution is
unstable. Equivalently, for a given temperature, whenever the mass is
below a certain value, the trivial solution is unstable.  In this
regime there is a new nontrivial static, stable solution whose
"non-normalizable" part vanishes. This means that in
the dual theory there is a new phase where the expectation value of
the operator dual to this scalar is non-zero, even in the absence of
any source. For any nonzero temperature there is a continuous phase
transition with mean field exponents at the critical mass. At zero
temperature (i.e. when the background is an extremal brane) the
transition persists, but is of the Berezinskii-Kosterlitz-Thouless
type. This setup is similar to that of holographic superconductors 
\cite{holosuper1, holosuper2,holosuper3} and
has been proposed as models for antiferromagentic transitions.

We consider quench across this critical point by working at the
critical mass, but turning on a time dependent source which crosses
zero (i.e. the critical point) at some time \cite{bdas}. We show that the dynamics
of the bulk scalar is dominated by a zero mode of the radial operator
in the critical region when the rate of change of the source is small,
This leads to a Landau-Ginsburg type dynamics with dynamical critical
exponent $z = 2$, and a resulting scaling behavior of the order
parameter.

\section{Probe Branes and Thermalization}
\vspace{0.5cm}
Probe branes in the bulk of $AdS$ have been used to introduce flavor
in the standard AdS/CFT correspondence. Consider for concreteness
$AdS_5 \times S^5$ whose dual is $\cN=4$ super-Yang-Mills in $3+1$
dimensions with gauge group $SU(N_c)$. Let us introduce $N_f$ Dp
branes which wrap a $AdS_m \times S^{p+1-m}$. Possible supersymmetric
wrappings are summarized in Table (\ref{tabl1}).

\begin{center}
\begin{table}[h]
\caption{\label{tabl1}Probe Branes in $AdS_5 \times S^5$.}
\centering
\begin{tabular}{|l|c|r|}
\hline
Brane&Wrapping&Dual Theory\\ \hline
D1& $AdS_2$ & $ 0+1 $ dim\\ \hline
D3& $AdS_3 \times S^1$ & $1+1$ dim\\ \hline
D5& $ AdS_4 \times S^2$ & $2+1$ dim\\ \hline
D7 & $AdS_5 \times S^3$ & $3+1$ dim \\ \hline
\end{tabular}
\end{table}
\end{center}

The Dp branes give rise to new hypermultiplet fields. These live on
the intersection of the Dp branes with the D3 branes which gave rise to
the $AdS_5 \times S^5$ geometry. From the point of view of the $N=4$
theory the hypermultiplets live on a lower dimensional defect. In the
strong coupling regime, the bulk theory is the original supergravity
together with the action of branes coupled to it.

In the limit of $N_f \ll N_c$ the backreaction of the probe branes on
the background $AdS_5 \times S^5$ geometry can be ignored and the
entire bulk theory is given by the action of these branes moving in
the fixed background geometry. We will take the brane action to be of
the DBI type. In the dual theory this means we can consider the defect
field theory by itself, and ignore the effect of hypermultiplet loops.

As is standard in the AdS/CFT correspondence, the boundary values of
the DBI fields are identified with sources for the dual operators in
the dual field theory.  Consider for example the case of a D5
brane. Let us write the $AdS_5 \times S^5$ metric in the form 
\ben
ds^2 =
(y^2+r^2)[-dt^2+dx_1^2+dx_2^2+dx_3^2]+
\frac{1}{y^2+r^2}[dr^2+r^2d\Omega_2^2+dy^2+y^2d(\Omega^\prime_2)^2]
\label{one}\een
The D5 brane is wrapped along $\xi^\alpha =
(t,r,\Omega_2,x_1,x_2)$. The fields in the DBI action are $y(\xi),
\Omega_2^\prime (\xi),x_3 (\xi)$, which are the transverse coordinates
to the brane. The value of $y(r=\infty)$ is then the mass of the
hypermultiplet fields coming from $(3,5)$ open strings joining the
D5 brane with the stack of $N_c$ three branes which produce the
background geometry. Thus a time dependent boundary value of $y$ is a
time dependent mass for the hypermultiplets.

Therefore a quantum quench in this dual theory may be implemented
simply by providing a time dependent boundary condition for the DBI
field. However the DBI fields are the transverse coordinates of the
branes - so this corresponds to a motion of the edge of the
brane. This disturbance sets up a wave along the brane and therefore
correponds to an excited state of the defect field theory. Our aim is
to figure out the nature of this state at late times.

In the full theory, such a disturbance would lead to a deformation of
the background geometry and possibly lead to black hole formation,
which would appear as thermalization in the boundary field theory. We
want to explore if any signature of thermalization remains in the
probe approximation. The following sections summarize some salient
points of work with Tatsuma Nishioka and Tadashi Takayanagi \cite{dnt}.

\subsection{Rotating D1 branes}
\vspace{0.5cm}

The essential physics is in fact apparent in the simplest example - D1
branes in $AdS_5 \times S^5$. For this purpose it is convenient to
write the $AdS_5 \times S^5$ metric as \ben ds^2 = 2drdv-f(r)dv^2+r^2
d\ts^2+(d\theta^2+\sin^2\theta d\vp^2+\cos^2\theta d\Omega^2_{3}) \een
If we are using the Poincare patch, $f(r) = r^2$ and $d\ts^2$ is the
flat metric on $R^3$, while in the global patch $f(r) = 1 +r^2$ and
$d\ts^2$ is the round metric on $S^3$. We have used
Eddington-Finkelstein coordinates in the $AdS_5$ part. The D1 brane is
along $(r,v)$ and its action is given by the standard DBI action
obtained from the induced metric. The dynamical fields on the brane
are $\theta (r,v), \vp (r,v), \Omega_3 (r,v)$ and the coordinates
contained in $d\ts^2$. It is clear from the symmetries that one can
have a class of solutions of the form \ben \vp (r,v),~~~~~~~~\theta =
\frac{\pi}{2} \een with all the other coordinates held constant. The
equations of motion which follow from the DBI action is best written
by using the advanced EF coordinate $u = v-2\int \frac{dr}{f(r)}$ as
well as $v$ \ben \partial_u\partial_v \vp + \frac{2}{L} \partial_v \vp
\partial_u \left( \frac{\partial_u \vp \partial_v \vp}{f(r)} \right) +
\frac{2}{L} \partial_u \vp \partial_v \left( \frac{\partial_u \vp
  \partial_v \vp}{f(r)} \right) = 0 \ ,
\label{uveqnmotion}
\een where \ben L = 1-\frac{4}{f(r)}\partial_u \vp \partial_v \vp \ .
\een It is easy to see that any $\vp$ which satisfies either
$\partial_u \vp = 0$ or $\partial_v \vp = 0$ is a solution of
(\ref{uveqnmotion}). In particular, a solution which is a function of
$v$ alone represents the retarded effect of a boundary value of
$\vp$.

The induced metric produced by such a retarded solution is given by
\be
\label{indf} ds_{ind}^2 = -f(r) dudv +(\partial_v \vp)^2 dv^2 =
2drdv - [ f(r)-(\partial_v \vp)^2]dv^2 \ .  \ee 
This is a
two-dimensional AdS Vaidya metric which has an
apparent horizon at $f(r) = (\partial_v \vp)^2$, provided this
equation has a solution for real $r$.  In the Poincare metric there is
always a solution for real $r$, while in global coordinates, this is
not guaranteed.

\begin{figure}[h]
\begin{minipage}{14pc}
\includegraphics[width=14pc]{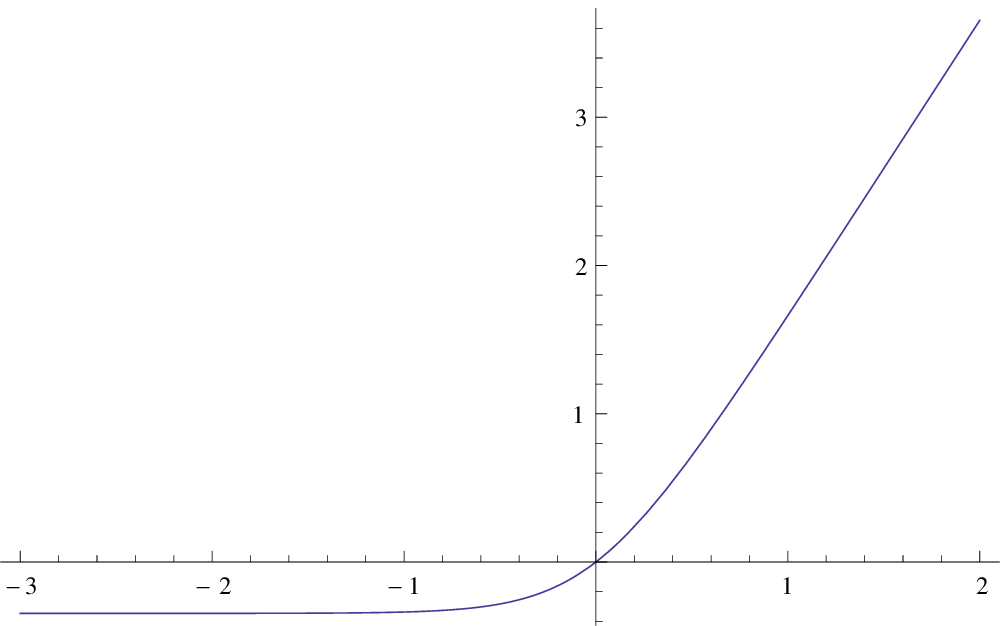}
\caption{\label{fig:prof1}The profile (\ref{prof1}) as a function of $v$}
\end{minipage}\hspace{2pc}%
\begin{minipage}{14pc}
\includegraphics[width=14pc]{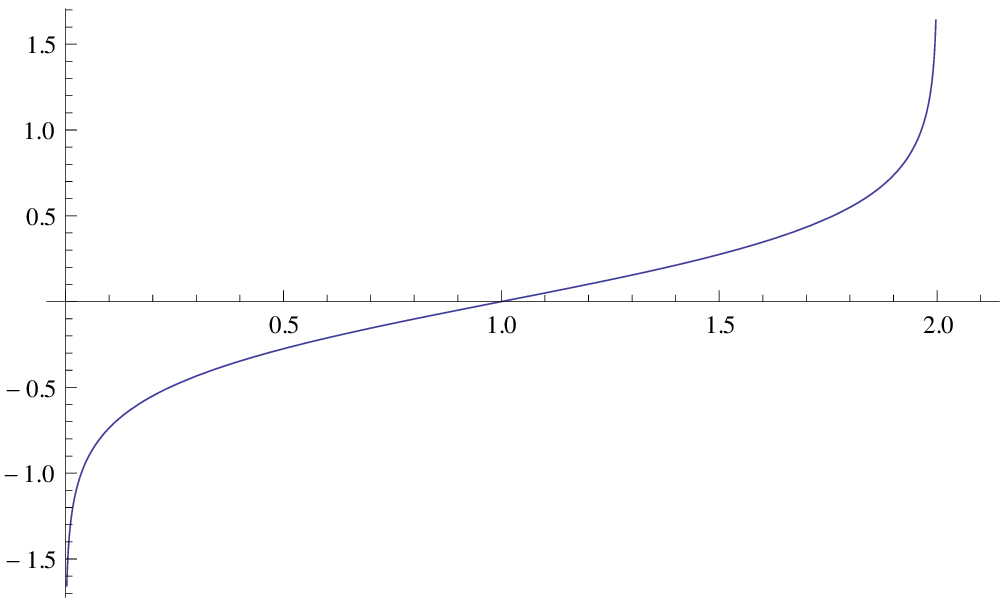}
\caption{\label{fig:ahprof1}Location of the apparent horizon. $v$
  as a function of $r$ from equation (\ref{ahprof1}). At late time this becomes an event horizon}
\end{minipage} 
\end{figure}

Depending on the profile of $\vp(v)$, the apparent horizon may or may
not develop into an event horizon. An example where it does is given
by the profile 
\ben 
\vp(v) = \vp_0 ( v + \frac{1}{k}\log \cosh (kv) )
\label{prof1}
\een 
which leads to the following equation for the location of the
apparent horizon for the Poincare patch 
\ben r = \vp_0 ( 1 + \tanh
(kv)) 
\label{ahprof1}
\een 
This asymptotes to an event horizon at $r = 2\vp_0$.  This
profile represents a D1 brane which starts from rest and spins with an
increasing spin, asymptoting to a constant rotation rate.  The
function (\ref{prof1}) and the location of the apparent horizon is
shown in Figure (\ref{fig:prof1}) and (\ref{fig:ahprof1})

\begin{figure}[h]
\begin{minipage}{14pc}
\includegraphics[width=14pc]{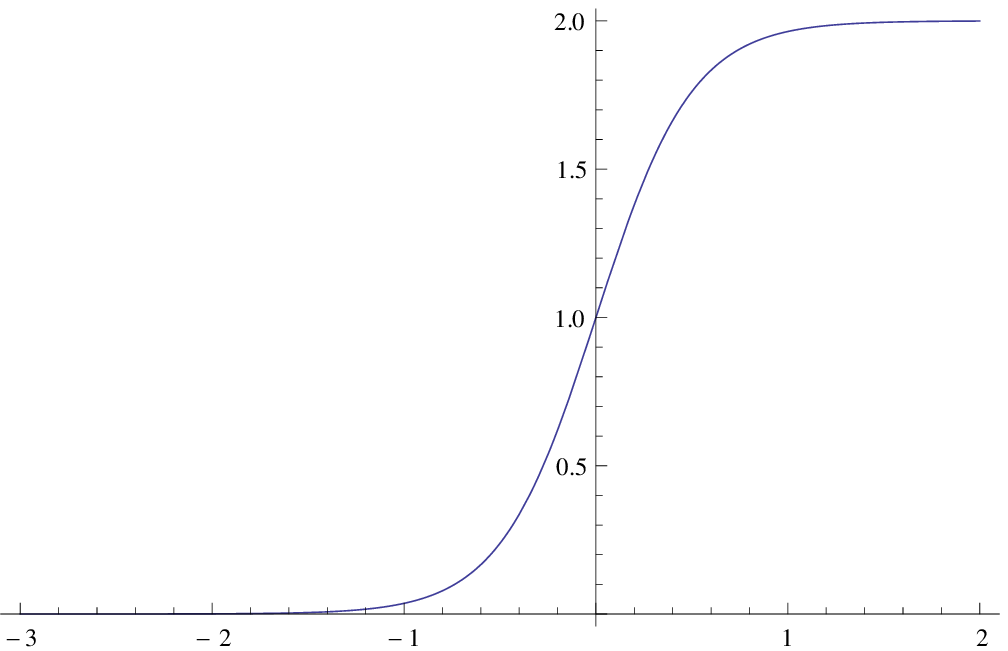}
\caption{\label{fig:prof2}The profile (\ref{prof2}) as a function of $v$}
\end{minipage}\hspace{2pc}%
\begin{minipage}{14pc}
\includegraphics[width=14pc]{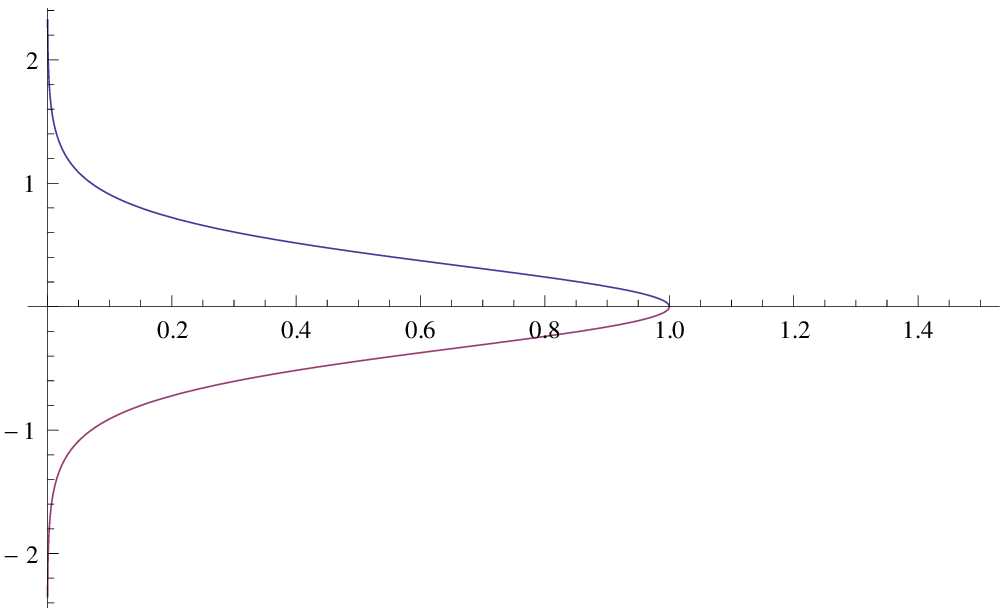}
\caption{\label{fig:ahprof2}Location of the apparent horizon. $v$
  as a function of $r$ from equation (\ref{ahprof2}). The apparent horizon now recedes back to $r=0$}
\end{minipage} 
\end{figure}

For strings
which eventually stop spinning, the apparent horizon does not develop
into an event horizon, but recedes back to $r=0$. For example if
\ben
\vp (v) = \vp_0(1 + \tanh(kv))
\label{prof2}
\een
the apparent horizon is located at 
\ben
r = \frac{k \vp_0}{\cosh^2 (kv)}
\label{ahprof2}
\een
 The
function (\ref{prof2}) and the location of the apparent horizon is
shown in Figure (\ref{fig:prof2}) and (\ref{fig:ahprof2})

In global $AdS$ the equation which determines the apparent horizon is
given by $1+r^2 = (\partial_v \vp)^2$ which does not always have a
solution for real $r$.  For example, a brane whose end point is uniformly rotating
has $\vp (v) = \omega v$ - this would lead to an event horizon only
when $\omega > 1$.

The Poincare patch solutions represent injection of energy from the
boundary, which flows into the Poincare horizon. In the global
solutions, the energy flows from a point of the boundary to the
antipodal point of the $S^3$.

Fluctuations of the brane around this classical solution will feel the
effect of an apparent horizon on the worldsheet.  Let us choose a static gauge
where the worldsheet coordinates are identified with two of the
space-time coordinates $\xi^a =
x^a,~~a=0,1$. The transverse coordinates are $x^I, I = 2 \cdots
9$. The $AdS \times S$ metric can be then written as \be ds^2 = g_{ab}
(x^a, x^I) dx^adx^b + G_{IJ}(x^a,x^K) dx^I dx^J \ .
\label{fluc1}
\ee
Expanding around a classical solution  $x_0^I(x^a)$,
\be
x^I(x^a) = x_0^I (x^a) + y^I(x^a) \ ,
\label{fluc3}
\ee
This leads to the following action for quadratic fluctuations
\be
S_2 = \frac{T_{D1}}{2}\int d^{2}\xi \sqrt{-\gamma_0} \gamma_0^{ab} G_{IJ}
(\xi^a,x_0^I) \partial_a y^I \partial_b y^J \ .
\label{fluc8}
\ee
where $\gamma_0^{ab}$ denotes the induced metric due to the background
solution $x_o^I$.
In particular, the fluctuations of $\varphi,\theta$ i.e. all fluctuations in $S^q$ directions
 are minimally coupled
massless scalars on the worldsheet, while the fluctuations of the
boundary gauge theory spatial directions $x^i, i
= 1\cdots 3$ have an additional factor of $r^2$ coming from the fact
that $G_{ij} = r^2 \delta_{ij}$ along these directions.

It is well known that fields which live on a space-time with an
apparent horizon behave approximately thermally if the apparent
horizon lasts long enough \cite{visser}. While the standard derivation of
Hawking radiation assumes the presence of an event horizon, the
essential physics is the large redshift, which is present near an
apparent horizon as well. In our case,  profiles like (\ref{prof1})
lead to exact thermality at late times since the apparent horizon
evolves into an event horizon. On the other hand profiles like
(\ref{prof2}) lead to an effective ``time dependent temperature'' in
the dual theory, which of course makes sense when the time variation
is slow enough.

The thermal nature of the state produced by time dependence becomes
clear from a calculation of the fluctuation
of the end-point of the string.
In \cite{Bra} it has been shown that the fluctuations
of a string suspended from the horizon of a AdS black brane ended at
a flavor D-brane near the boundary of AdS are dual to Brownian
motion of the corresponding quark in the hot $\CN = 4$ gauge
theory. In this case the bulk black brane metric induces a worldsheet
metric which has a horizon. The fluctuations then reflect Hawking
radiation from the worldsheet horizon. 

In the D-brane solutions considered above, the bulk metric has no
horizon. However due to the motion of the D-brane, the induced metric
on the worldvolume can develop a horizon. Since the fluctuations of
\cite{Bra} comes purely from properties of the induced metric it
is natural to expect that a similar phenomenon appears in our case.

The result of this calculation for fluctuations in the $\vp$
direction is 
\ba \langle (\Delta y^\varphi (t-t^\prime))^2\rangle &
\sim & \frac{\pi (t-t^\prime)^2}{12 \beta^2} \ ,~~~~~~~~~~~~~~~~~~~\pi
(t-t^\prime) \ll \beta \ ,\no \langle (\Delta y^\varphi
(t-t^\prime))^2\rangle & \sim & \frac{(t-t^\prime)}{2 \beta}
-\frac{1}{2\pi} \log [2\pi (t-t^\prime)/\beta ] \ ,~~~~~\pi
(t-t^\prime) \gg \beta \ , \ea while for fluctuations in the $\vx$
direction are \ba \langle [\Delta y^i (t-t^\prime)]^2 \rangle & \sim &
\frac{(t-t^\prime)^2}{m\beta} \ ,~~~~t \ll m\beta^2 \ ,\no & \sim &
\beta |t-t^\prime| \ ,~~~~t \gg m\beta^2 \ .  \ea

The rotating $D1$ brane solution corresponds to a time dependent
coupling in the $N=4$ theory coupled to hypermultiplets living on the
zero dimensional defect.  The D1-D3 system is 1/4-BPS and the D1-D3
open strings lead to the two complex scalars $(Q,\ti{Q})$ of
hypermultiplets which belong to the fundamental and anti-fundamental
representations of the color $SU(N)$ gauge group. Let us
express the three complex adjoint scalar fields
in the $\CN=4$ super Yang-Mills by $(\Phi_1,\Phi_2,\Phi_3)$. These
correspond to cartesian coordinates in the transverse $C^3$ composed of
$(r,\Omega_5)$ where $\Omega_5$ represents the 5-sphere.  We choose
$\Phi_3$ such that its phase rotation describes the one in the $\vp$
direction and that $\theta=\pi/2$ is equivalent to
$\Phi_1=\Phi_2=0$.  The time dependent coupling term
which corresponds to a uniformly rotation D1-brane is given by
\be \int dt
\left[\ov{Q}~ \left[\mbox{Im}(\Phi_3 e^{-i\omega t})\right]^2~Q
  +\ti{Q}~\left[ \mbox{Im}(\Phi_3 e^{-i\omega
      t})\right]^2~\ov{\ti{Q}}\right] .
\label{intti}
\ee
The justification for this is given in \cite{dnt}.
For non-uniform rotation with a profile $\vp (v)$ the exponential
factors are simply replaced by $e^{\pm i \vp(t)}$.

Thus, from the boundary theory point of view we have a time dependent
coupling - this leads to thermalization. Note that this is
thermalization of only the hypermultiplet sector - the vector
multiplet sector is unchanged in the lowest order of this
approximation. 

\subsection{Higher dimensional probes}
\vspace{0.5cm}

Thermalization in higher dimensional field theories can be also
investigated by considering higher dimensional branes in the bulk of
$AdS \times S$.  It is possible to construct uniformly rotating D7 and
D5 branes using a combination of analytic and numerical methods.
While these D5 and D7 solutions have been obtained earlier in
\cite{evans}, 
the implications to thermalization was not realized.
Quench-like solutions (i.e. solutions where the rotation vanishes in the
asymptotic past and the asymptotic future) can be obtained numerically
as well \cite{dnt}.

Rotating D5 branes are dual to a time dependent mass of the
hypermultiplets in the dual 2+1 dimensional defect field
theory. Rotating D7 branes are dual to a time dependent phase of the
mass of fermions in the hypermultiplet, as well as a time dependent
bosonic potential. The essential physics is similar to the D1 brane
descibed in the previous subsection, viz. an apparent horizon is
formed and the fluctuations respond in a thermal fashion.

\begin{figure}[h]
\begin{minipage}{14pc}
\includegraphics[width=14pc]{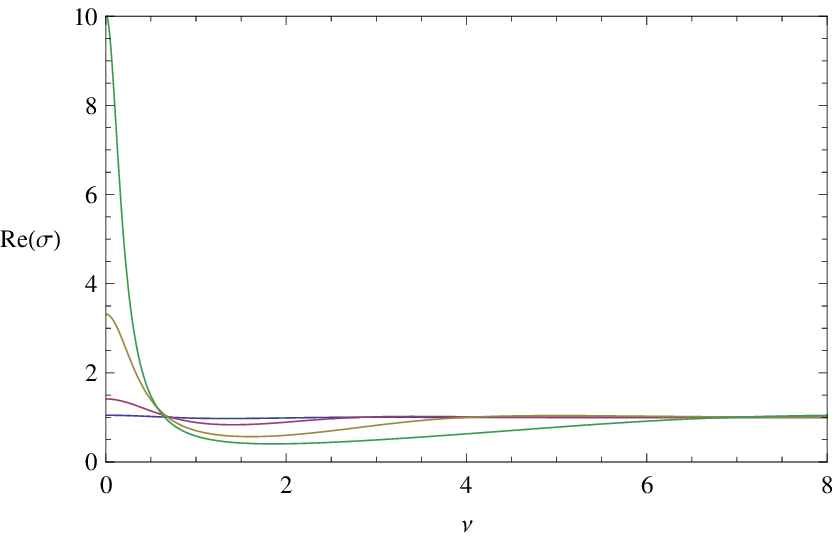}
\caption{\label{fig:one}Real part of $\sigma(\nu)$.}
\end{minipage}\hspace{2pc}%
\begin{minipage}{14pc}
\includegraphics[width=14pc]{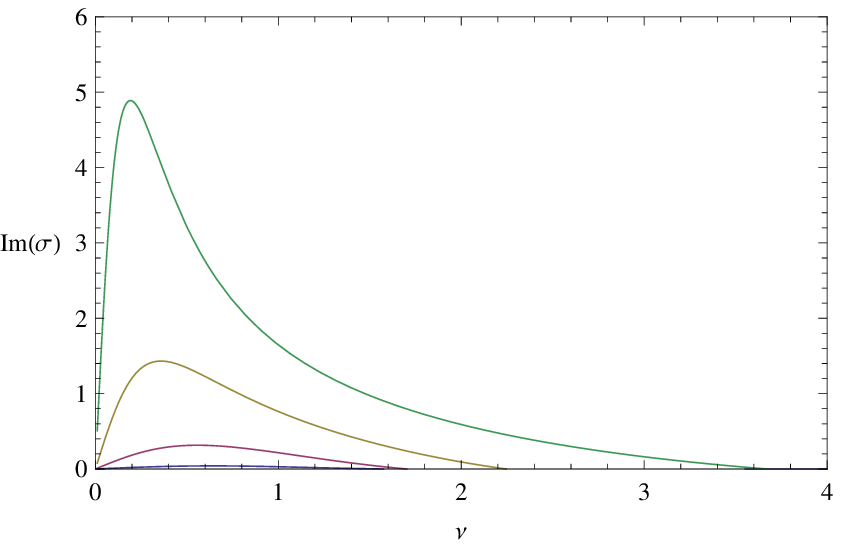}
\caption{\label{fig:two}Imaginary part of $\sigma(\nu)$.}
\end{minipage} 
\end{figure}

An additional signature of dissipation appears when we consider probe
D3 branes obtained by performing T-duality on the D1 brane solution
described above along $x_1$ and $x_2$ directions.  In this case the
worldvolume is a $3+1$ dimensional theory and the dual defect field
theory is $2+1$ dimensions. It turns out that in this case one can
turn on a background electric field on the worldvolume of a uniformly
rotating D3 brane. This means we are turning on a chemical potential
and a charge density for the corresponding global charge in the dual
field theory, in addition to a time dependent coupling. The
fluctuations of the gauge field around this background now react to
the apparent horizon of the induced metric. This leads to an
electrical conductivity $\sigma(\nu)$ , whose behavior as a function
of the frequency $\nu$ is quite similar to Drude theory at low
frequencies. However the real part of $\sigma (\nu)$ approaches a
constant at large frequency, as is typical in a 2+1 dimensional
critical theories. This is shown in Figures (\ref{fig:one}) and
(\ref{fig:two})

\subsection{Time Dependent Chemical Potential}
\vspace{0.5cm}
An interesting example of the process of thermalization due to
formation of an apparent horizon on the worldvolume concerns the
thermalization of the meson sector of $N=4$ Yang-Mills theory due to a
time dependent chemical potential \cite{hashimoto}. As usual, quarks
are introduced by placing a set of D7 branes and a chemical potential
corresponds to a worldvolume electric field. This is made time
dependent by coupling to a time dependent external current, i.e. by
injecting quarks from outside. As in the previous examples this
results in the formation of an apparent horizon with a characteristic
temperature. For other effects of this phenomenon see
\cite{otherapparent}.

Finally the formation of apparent horizons in these examples is similar to acceleration horizons on worldvolumes discussed in \cite{wa} and related phenomena have been studied in \cite{Gursoy}.

\section{Quench across a Holographic Critical Point}
\vspace{0.5cm}
As we discussed above, quantum quench is particularly interesting when
the time dependent coupling crosses a critical point.  
The probe
approximation has turned out to be quite useful for studying
holographic critical points. Such phase transitions are known to occur
for various probe branes \cite{phase,karch,bktpapers}.

Another class of probe fields appear in discussions of holographic
superconductors \cite{holosuper1,holosuper2,holosuper3}. 
This setup consists of a charged scalar field in the
presence of a charged black brane. When the gauge coupling is large,
the backreaction of the scalar and the gauge field to the background
geometry can be ignored. In this case, for a given mass of the scalar
field there is a critical temperature below which the scalar condenses
- this is interpreted as superfluidity in the boundary theory. The
phase transition between the ordered and disordered phases is a
critical point.

An even simpler setting consists of a neutral scalar field with a
quartic coupling which is large enough to ensure that a probe
approximation is reliable. As shown in \cite{liu1} for suitable values
of the parameters, this model displays a critical point of the type
encountered in antiferromagentic phase transitions. In this section I
will describe a recent study of quantum quench in this model in
collaboration with Pallab Basu
\cite{bdas}. 

\subsection{The equlibrium phase transition}
\vspace{0.5cm}
The model of \cite{liu1} has a neutral scalar field $\phi(t,r,\vx)$ in
the background of a charged $AdS_4$ black brane. The lagrangian is
given by 
\ben 
\cL =
\frac{1}{2\kappa^2\lambda}\sqrt{-g}[-\frac{1}{2}(\partial \phi)^2
  -\frac{1}{4}(\phi^2+m^2)^2-\frac{m^4}{4}]
\label{1-1}
\een
The background metric is given by (in $R_{AdS}=1$ units)
\ben
ds^2 = [-r^2 f(r)dt^2+r^2 d\vx^2]+\frac{dr^2}{r^2 f(r)}
\label{1-2}
\een
where
\ben
f(r) = [1+\frac{3\eta r_0^4}{r^4}-\frac{(1+3\eta) r_0^3}{r^3}]~~~~~~~~0 \leq \eta \leq 1
\label{1-3}
\een
The associated Hawking temperature is then given by
\ben
T = \frac{3}{4\pi r_0}(1-\eta)
\label{1-4}
\een
In the following we will replace $r \rightarrow r r_0$. This means all dimensional quantities are expressed in units of $r_0$.

In the limit of large $\lambda$ the
field $\phi$ can be regarded as a probe field.
In \cite{liu1} it was shown that when the mass lies in the range 
\ben
-\frac{9}{4} < m^2 < -\frac{3}{2}
\label{1-5} 
\een 
there is a critical phase transition at
some value of $T = T_c(m)$ when the source to the dual operator
vanishes.  
Conversely, for a given $T$ there is a
value of $m^2 = m_c^2$ where the theory is critical.

The upper limit in (\ref{1-5}) is the BF
bound for the near-horizon $AdS_2$ geometry which appears in the
extremal ($\eta = 0$) metric. (Note that the AdS scale for this
infrared $AdS_2$ is given by $1/\sqrt{6}$ in our units). The lower
bound is the BF bound for the asymptotic $AdS_4$.  Field
configurations which are translationally invariant in the $\vx$
directions satisfy the equations of motion 
\ben
\frac{1}{r^2}[-\frac{1}{f(r)}\partial_t^2+\partial_r(r^4
  f(r)\partial_r)]\phi -m^2 \phi -\phi^3 = 0 
\label{1-6}
\een 
Near the $AdS_4$
boundary the asymptotic behavior of the solution to the linearized
equation is of the form \footnote{When we turn on $J(t)$, it is a valid concern whether we will be able to neglect the non-linear term near the boundary. This can be done as long as $\Delta>0$ or $m^2<0$.}
\ben
\phi (r) = J(t)r^{-\Delta_-}[1+O(1/r^2)]+<{\cal O}>(t) r^{-\Delta_+}[1+O(1/ r^2)]
\label{1-7}
\een
where $\Delta$ is given by
\ben
\Delta_\pm = \frac{3}{2} \pm \sqrt{m^2 + \frac{9}{4}}
\label{1-8}
\een
In the range of masses of interest, both the solutions are
normalizable, so that there is a choice of quantization. The standard
quantization considers the coefficient $J(t)$ as the source in the
dual field theory and $B(t)$ then gives the expectation value of the
dual operator. In the alternative quantization the expectation and source change the role.

Consider first the linearized problem, ignoring the cubic term. By a standard change of coordinates to tortoise coordinates $\rho$ and a field redefinition to 
$\chi$,
\ben
d\rho =- \frac{dr}{r^2 f(r)}~~~~~~~~~~~~~~\phi (r,t) = \frac{\chi (\rho,t)}{r}    
\label{1-9}
\een
The horizon is then at $\rho = \infty$ and the boundary is at $\rho = 0$. 
At the linearized level, the equation (\ref{1-6}) becomes 
\ben
-\partial_t^2 \chi = -\partial_\rho^2 \chi + V_0(\rho) \chi \equiv \cP_\rho \chi
\label{1-10}
\een
with
\ben
V_0(\rho) = r^2 f(r)[(m^2+2) - \frac{6\eta}{r^4}+\frac{1+3\eta}{r^3}]
\label{1-11}
\een
where in $V_0(\rho)$ we need to express $r$ in terms of $\rho$ using
(\ref{1-9}).  

\begin{figure}[h]
\begin{center}
\includegraphics[scale=0.75]{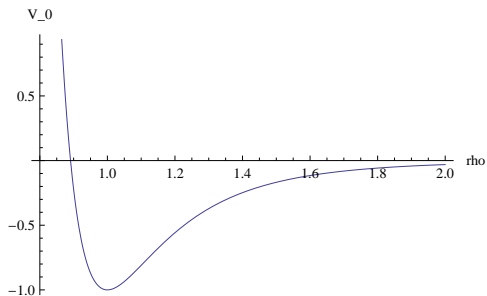}
\end{center}
 \caption{\label{fig:potential1} 
The potential $V_0(\rho)$}
\end{figure}

For solutions of the type $\chi \sim e^{-i\omega t}$, equation
(\ref{1-11}) is a Schrodinger problem in a potential $V_0(\rho)$. The
potential goes to zero at the horizon $\rho = \infty$ and behaves as
$\frac{(m^2+2)}{\rho^2}$ near the boundary $\rho = 0$.  Note that for
a brane background at any finite temperature, $f(r) \sim (r-1)$ near
the horizon, while $\rho \sim -\log (r-1)$ so that $V_0 \sim
e^{-\rho}$ as we approach the horizon. A typical $V_0(\rho)$ for $m^2
> -2$ is shown in Figure(\ref{fig:potential1}).
In contrast, for the
extremal background $f(r) \sim (r-1)^2$ while $\rho \sim 1/(r-1)$ so
that $V_0 \sim 1/\rho^2$. This makes the analysis for the extremal
background rather subtle.  We will work with the
non-extremal case.

In \cite{liu1} it was shown that when $m^2$ is below a critical value,
$m_c^2$ there are bound states of this Schrodinger problem, showing
that the trivial $\phi=0$ is unstable.  At $m^2 = m_c^2$ a zero energy
bound state appears, which vanishes in an appropriate fashion at the
boundary and is in addition regular at the horizon.  In the
complex frequency plane some quasinormal mode(s) hit the origin at
$m^2 = m_c^2$. This critical mass is $m_c^2 = -\frac{3}{2}$ when $\eta
= 1$ and decreases with decreasing $\eta$ or increasing temperature.

For $m^2 < m_c^2$ there is a stable nontrivial static solution $\phi_0
(r)$ of the full nonlinear equation of motion with the condition that
$J = 0$.  This means that in the conventional quantization, the
expectation value of the dual operator is nonzero even in the absence
of a source, i.e. the field
condenses. The the critical point is at $m^2 =m_c^2$ and $J = 0$.
Similarly there is a nontrivial solution with $B = 0$ which means that
there is a condensate in the alternative quantization as well.

The critical point has mean field exponents at any finite
$T$. If the operator dual to the field $\phi$ is $\cO$ and the source
is $J$ then an analysis identical to that presented in \cite{liu1}
leads to (for $m^2-m_c^2 \rightarrow 0^+$) \ben <\cO>_{J=0}\sim
(m^2-m_c^2)^{1/2}~~~~~\frac{d<\cO>}{dJ}|_{J=0}\sim
(m^2-m_c^2)^{-1}~~~~~~<\cO>_{m=m_c}\sim J^{1/3}
\label{1-12}
\een 
Exactly at zero temperature the phase transition is of BKT type
and the order parameter depends exponentially 
\ben <\cO>_{J=0} \sim
\exp \left[-\frac{\pi\sqrt{6}}{2\sqrt{m_c^2-m^2}}\right]
\label{1-13}
\een

\subsection{Quenching across the Critical Point}
\vspace{0.5cm}
In the following, we will study quench across this critical point by considering a time dependent source $J(t)$ which asymptotes to constant values at early and late times and crosses zero at some time, e.g.
\ben
J(t) = J_0 \tanh (vt)
\een
In the conventional quantization this means that in the dual boundary field theory, we have a source coupling to the operator dual to $\phi$. For any static $J \neq 0$ we of course have a nontrivial $\phi(r)$ and hence a nonzero $<\cO>$. Our first aim is to get some insight into the time dependence of $<\cO (t)>$ when we have a nontrivial $J(t)$.

\subsection{Breakdown of Adiabaticity}
\vspace{0.5cm}
If the time dependence is slow enough one would expect that far away from the critical point the dynamics is adiabatic, while near the critical point adiabaticity should break down. It is instructive to examine the way this happens. 

It is well known that to study low frequency modes in the background
of a black brane it is convenient to use ingoing Edddington-Finkelstein
coordinates,  
\ben
u=t- \rho,~~~~~~\rho
\een
where $\rho$ is defined in (\ref{1-9}).
In terms of these coordinates the equation of motion \ref{1-6}) becomes
\ben
-2\partial_u\partial_\rho \chi = -\partial_\rho^2 \chi + V(\rho,\chi).
\label{4-1}
\een
where
\ben
V(\rho,\chi)= V_0(\rho)\chi+ f(r) \chi^3.
\een
This equation has to be solved with the boundary condition that the
field is {\em regular at the horizon}, which at the linearized level
is equivalent to requiring that the waves are purely ingoing at the
horizon \cite{bhmr,hubeny2}. 
We need to solve (\ref{4-1}) with the condition
\ben
\chi(u,\rho) \rightarrow \rho^{-1+\Delta_-} J(u) ~~~~~~{\rm as}~~ \rho
\rightarrow 0
\label{4-1-1}
\een
where $\Delta_\pm$ are defined in (\ref{1-8}).
To perform the adiabatic expansion, let us decompose the field $\chi
(\rho, u)$ as
\ben
 \chi(\rho,u)=\chi_l(\rho,u)+\chi_s(\rho,u) 
\label{4-1-2}
\een
Where $\chi_l(\rho,u)=J(u) \rho^{-1+\Delta_-}$ and $\chi_s(\rho,u)
\sim \rho^{-1+\Delta_+}$ as $\rho \rightarrow 0$.  

For a constant $J$,  $\chi_l(\rho,u)=\chi_l(\rho)$ is time independent.
In this case there is a static
solution $\chi_s(\rho,u)=\chi_0(\rho)$, which is the equlibrium configuration. In the presence of a source which is {\em slowly varying} in units of the horizon radius, one can therefore expand the field $\chi_s(\rho,u)$ in an adiabatic expansion of the form
\ben \chi_s(\rho,u) = \chi_0(\rho, J(u)) +
\epsilon ~\chi_1(\rho,u) + \cdots.  
\label{expan}
\een 
Here $\epsilon \sim \partial_u$  (recall that we are using $r_0=1$ units) is an adiabaticity parameter which keeps track of the adiabatic
expansion. If we scale $u \rightarrow u/\epsilon$, each $u$ derivative
is of order $O(\epsilon)$. The idea then is to insert (\ref{expan})
into the equations of motion and obtain equations for $\chi_1,
\chi_2,\cdots$ order by order in $\epsilon$. To the lowest order one
gets 
\ben
\cD_\rho^{(1)} \chi_1 =   \{  [-\partial_\rho^2+ V_0(\rho)]  +f(r)
    (3\chi^2_0+6 \chi_l\chi_0+3\chi_l^2) \} \chi_1= - 2
    \partial_u\partial_\rho \chi_l - 2 \partial_u\partial_\rho \chi_0
\label{1-16} 
   \een The solution to this equation is 
\ben 
\chi_1=\int d\rho'
   G(\rho,\rho')\partial_{u^\prime}\partial_{\rho^\prime}
   (\chi_0+\chi_l)(\rho^\prime).
\label{1-17a}
\een 
where $G(\rho,\rho^\prime)$ is the Green's function of the
operator $\cD_\rho^{(1)}$ with the boundary conditions $G(0,\rho) =0$
and $G(\infty,\rho)$ is regular : 
\bea G^{(1)}(\rho,\rho') & = &
\frac{1}{W(\ttxi_1,\ttxi_2)} \, \ttxi_1(\rho') \ttxi_2(\rho) , \quad
\rho < \rho' \nn \\ & = & \frac{1}{W(\ttxi_1,\ttxi_2)} \,
\ttxi_2(\rho') \ttxi_1(\rho), \quad \rho > \rho', 
\eea where $\ttxi_1$
and $\ttxi_2$ are solutions of homogeneous part of eqn (\ref{1-17a})
satisfying appropriate boundary condition at the horizon $\rho
=\infty$ and the boundary $\rho = 0$ respectively, and
$W(\ttxi_1,\ttxi_2)$ is the Wronskian which is independent of $\rho$
in this case. We have normalized $\ttxi_1(\rho)$ and $\ttxi_2(\rho)$
in such a fashion that $\ttxi_1=1$ at the horizon and $\ttxi_2
\rightarrow \rho^{-1+\Delta_-}$ near the boundary. Regularity of the
functions $\chi_l$ and $\chi_0$ mean that $\partial_r \chi_l,
\partial_r \chi_0$ are finite at the horizon.  Since $(r-1) \sim
e^{-\rho}$, this implies that $\partial_\rho(\chi_l+\chi_0) \sim
\exp(-\rho)$. This ensures that the integral in (\ref{1-17a}) is
finite even though the Green's function approaches a constant in the
region near the horizon $\rho^\prime \rightarrow \infty$.  Furthermore, near the
horizon $\ttxi_2$ can be expressed as a linear combination of a
regular and irregular solution, i.e $\ttxi_2(\rho \rightarrow
\infty)=a \rho + b$. This implies that $W(\ttxi_1,\ttxi_2)
=a$.  Thus $\chi_1(u,\rho)$ is finite so long as $a$ is finite.

At the critical point, $J$ becomes small. Then $\chi_0$ and $\chi_l$ in the left hand
side of (\ref{1-16}) vanish, and the operator is identical to the
operator acting on the linearized small fluctuations at $m^2 = m_c^2$
around the trivial solution $\chi_0 = 0$, i.e. the operator $\cP_\rho$
which
appears on the right hand side of (\ref{1-10}). We know that this
operator has a zero mode which is regular at the horizon {\em and}
vanishes as $\rho^{-1+\Delta_-}$ at the boundary $\rho = 0$. This
means that at this point $a = 0$. Therefore, the first adiabatic
correction diverges.
 For small
$J(u)$, the leading departure from the critical operator comes from
the term which is proportional to $\chi_0^2 \sim J^{2/3}$.  Thus we
can use perturbation theory in $J$ to estimate $a \propto J^{2/3}$. As
argued before, $\chi_0 \sim (-J)^\frac{1}{3}$, while $\chi_l \sim J$.
Hence the leading divergence in $\chi_1$ can be estimated as 
$\chi_1(\rho,u) \sim \ J^{-4/3}\dot{J}$. Adiabaticity breaks down when 
\ben
\chi_1(\rho,u) \sim \chi_0  \Rightarrow \dot{J} \sim J^{5/3}
\een
In particular for profiles of $J(u)$ where $J(u) \sim vu$ near the critical point at $J=0$, adiabaticity breakdown occurs at 
\ben
u \sim v^{-2/5}
\een

\subsection{Quench in a Landau Ginsburg Model}
\vspace{0.5cm}
The scaling behavior found above is identical to that in a
Landau-Ginsburg dynamics with dynamical critical exponent $z=2$. The
dynamics of an spatially homogeneous order parameter $\vp$ is given by
\ben \frac{d\vp}{dt} +m^2 \vp + \vp^3+J(t) = 0
\label{three}
\een
The equilibrium critical point is at $m = J = 0$. For $m^2=0$ the
 equilibrium value of the order parameter is 
\ben
\vp_0(J) = [-J]^{1/3}
\een
As usual an adiabatic expansion is of the form
\ben
\vp (t) = \vp_0 (J(t)) + \epsilon \vp_1 (t) + \cdots
\een
and to lowest order
\ben
\vp_1 =\frac{1}{2\vp_0^2} {\dot{J}}\frac{\partial \vp_0}{\partial J}
\een
and adiabaticity breaks down when $\vp_1 \sim \vp_0$ which becomes the condition 
\ben
{\dot{J}}\frac{\partial \vp_0}{\partial J} \sim J^{1/3} \Rightarrow {\dot{J}} \sim J^{5/3}
\een
exactly as in our system. 

\begin{figure}[h]
\begin{center}
\includegraphics[scale=0.75]{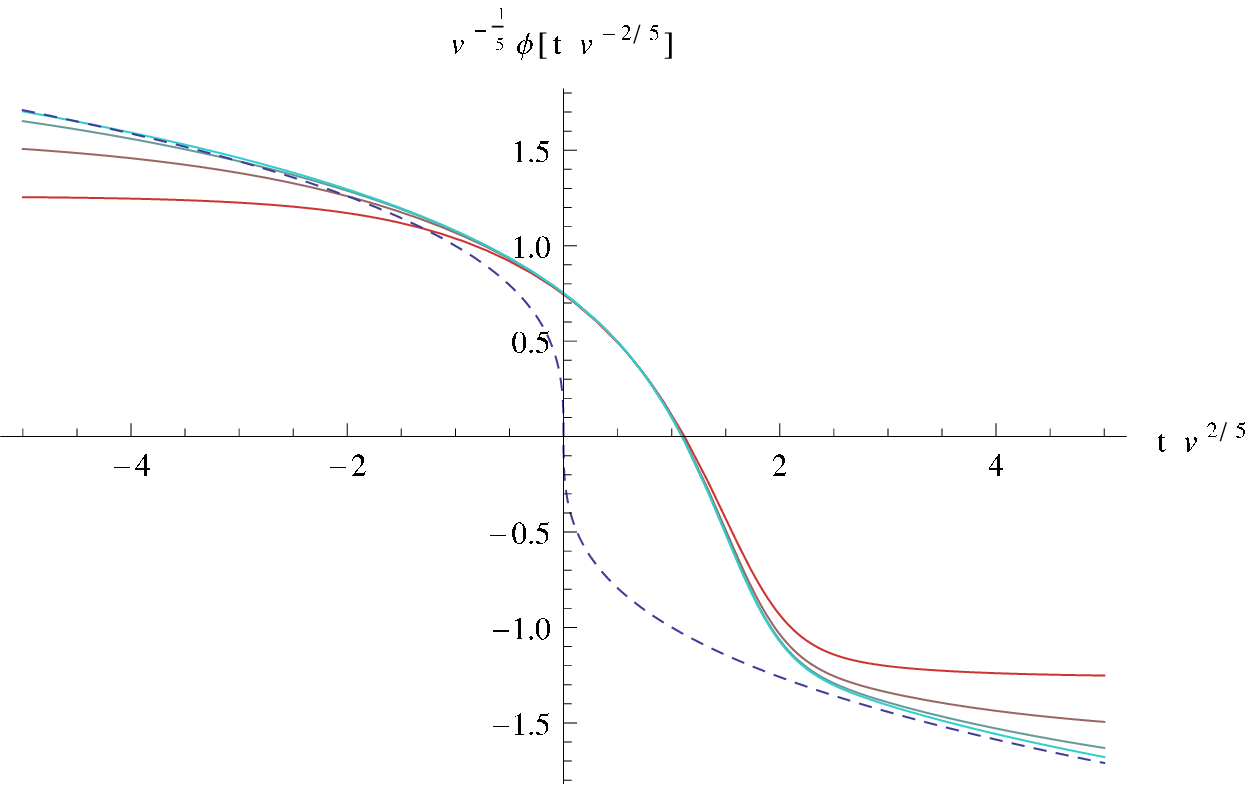}
\end{center}
 \caption{\label{fig:LG} 
The scaled order parameter as a function of scaled time for a $J(t) =
   \tanh(vt)$ at $m^2=0$ with $v=10^{-0.5},10^{-1},10^{-1.5},10^{-2}$
   (from the bottom on the left). The adiabatic solution (dashed) is
   also shown as a comparison.}
\end{figure}

When adiabaticity breaks down our system enters a scaling
region. Suppose the function $J(t)$ behaves linearly with time in the
critical region. Then it is straightforward to see from (\ref{three})
that in this region the solution is of the form \ben \vp (t,v) =
v^{1/5} \vp(tv^{2/5},1) \een This means, in particular, that the time
at which the order parameter hits zero scales as $v^{-2/5}$ while the
value of the order parameter at $t=0$ scales as $v^{1/5}$. A numerical
solution of the equation (\ref{three}) with adiabatic initial
conditions is consistent with this scaling, as shown in Figure (\ref{fig:LG})

Note that the order parameter hits zero {\em later} than the location
of the equilibrium critical point. This is a manifestation of the
phenomenon of raising the critical temperature when the temperature is
time dependent \cite{bao1} which has been holographically realized in
\cite{bao2}. 

We will now argue that the behavior of our holographic system in the
critical region is fairly well described by such a LG dynamics

\subsection{Small $v$ dynamics in the Holographic Model}
\vspace{0.5cm}
Consider the dynamics of the bulk field in the critical region in
the presence of a linear quench $J(u) = vu$ for small $v$. To do this,
first substitute (\ref{4-1-2}) in the equation (\ref{4-1}) and rescale
\ben 
\chi_s \rightarrow v^\frac{1}{5} \tilde \chi_s,u \rightarrow
v^{-\frac{2}{5}}\tilde u 
\label{rescale}
\een 
The equation (\ref{4-1}) then becomes
\ben [-\partial_\rho^2+ V_0(\rho)] \tch_s+v^{\frac{2}{5}} [f(r)
  (\tch_s)^3 + \tilde u [-\partial_\rho^2+ V_0(\rho)] \tch_l+2
  \partial_{\tilde u}\partial_{\rho} \tch_s ] +\cdots=0
\label{veq}
\een
The ellipsis denote terms which contains higher powers of $v$.

Let us expand the sub-leading part of the scalar field in terms of eigenfunctions of the operator $\cP_\rho$ (defined in equation (\ref{1-10}) at  the critical point,
\ben
\tch_s(\rho,u) = \int \tilde a_k(u) \ch_k(\rho)  dk 
\een
The $\ch_k$ satisfy
\ben
\cP_\rho^c \chi_k = [-\partial_\rho^2+ V_0^c(\rho)]\chi_k = k^2 \chi_k
\label{eigeneqn}
\een where $V_0^c$ denotes the potential in (\ref{1-11}) at
$m^2=m_c^2$.  The eigenfunctions $\chi_k(\rho)$ are delta function
normalized and obey the condition 
\ben 
{\rm Lim}_{\rho \rightarrow 0}
[ \rho^{1-\Delta_-} \chi_k (\rho) ] = 0 \een In terms of the
eigen-coefficients $a_k(u)$ the equation (\ref{veq}) becomes, 
\ben 
k^2
\tilde a_k+ v^{\frac{2}{5}}\left( \tilde u {\cal J}_{k} + \int b_{kk'}
\partial_{\tilde u} \tilde a_{k'} dk' +  \int \tilde a_{k'} \tilde
a_{k''} \tilde a_{k'''} C_{k,k',k'',k'''} d k' dk'' dk''' \right) +\cdots =0
\label{modeeqn}
\een where 
\bea {\cal J}_{k}& = &\int \ch_k (\rho) [-\partial_\rho^2+
  V_0(\rho)] \ch_l \, d\rho \nn \\ 
b_{kk'} & = & \int d \rho \, \ch_k
\partial_\rho \ch_k' \nn \\ 
C_{k,k',k'',k'''}& = &\int d\rho \, \ch_k
\ch_{k'} \ch_{k''} \ch_{k'''} f(r) . 
\label{moments} 
\eea 
The equation
(\ref{modeeqn}) suggests that there is a solution in a perturbation
expansion of powers of $v^\frac{2}{5}$, 
\ben  \tilde
a_k(\tilde u) =\delta(k) \tilde \xi_0(\tilde u) + v^{\frac{2}{5}}
\tilde \eta_k(\tilde u)+\cdots, 
\label{vexpn}
\een 
If this expansion makes sense, the dominant behavior of the solution for $\chi_s$ is given by the zero mode $\tilde\xi_0$.

Substituting (\ref{vexpn}) in the equation (\ref{modeeqn}) we get to
the lowest order in the small $v$ expansion,
\bea &
& \tilde u \, {\cal J}_{0} + b_{00} \frac{d}{d\tilde u} \tilde
\xi_0(\tilde u) + C_{0000} \tilde \xi_0(\tilde u)^3 = 0 \nn \\ & &
\tilde \eta_k(\tilde u) = -\frac{1}{k^2}\left (\tilde u {\cal J}_{k}-
b_{k0} \frac{d}{d\tilde u} \tilde \xi_0(\tilde u)-C_{k000} \tilde
\xi_0(\tilde u)^3\right)
\label{scalingeqn}
\eea
The first equation in (\ref{scalingeqn}) determines the time
dependence of the zero mode $\tilde\xi_0$, while the second equation
determines the leading correction from nonzero modes in terms of the
solution for $\tilde\xi_0$.

It is useful to rewrite the second equation by subtracting the first from it,
\ben
\tilde \eta_k(\tilde u) = -\frac{1}{k^2}\left ( \tilde u ({\cal J}_{k}-{\cal J}_{0}) -
 (b_{k0}-b_{00} )\frac{d}{d\tilde u} \tilde \xi_0(\tilde
 u)-(C_{k000}-C_{0000}) \tilde \xi_0(\tilde u)^3\right)
\label{scalingeqn2}
\een
The expansion in powers of $v^{2/5}$ would be valid if $\eta_k(\tilde u)$ remains finite.
However, $k$ is a continuous parameter starting from zero. This means that there is a potential divergence in the $k \rightarrow 0$ limit.

Indeed, as will be argued in the next subsection, for generic
potential $V_0$ the numerator on the right hand side of
(\ref{scalingeqn2}) behaves as $k$ for small $k$, so that $\eta_k$
indeed diverges at $k=0$. However exactly at the critical point, the
small $k$ behavior changes to $k^2$ so that $\eta_k$ remains finite
and the expansion in $v^{2/5}$ remains valid.

\subsection{Validity of the small $v$ expansion}
\vspace{0.5cm}
To examine this issue we need to consider the eigenvalue
problem \ben [-\partial_\rho^2+ V_0(\rho)]\chi_k = k^2 \chi_k
\label{egen}
\een 
As discussed above the potential $V_0(\rho) \rightarrow
-e^{-\rho}$ as $\rho \rightarrow \infty$. This potential is shown in Figure (\ref{fig:potential1}).

The basic features of the eigenfunctions can in fact be gleaned from a
simpler problem in which we replace the potential by the following
potential which has the same qualitative features.  \bea & = &
\infty~~~~~~~~~~\rho = 0 \nn \\ U(\rho) & = & -U_0~~~~~~~~~~~~0 \leq
\rho \leq 1 \nn \\ & = & 0~~~~~~~~~~~~~~~1 \leq \rho \leq \infty
\label{squarewell}
\eea
This problem is of course solvable. The eigenfunctions of the
Schrodinger operator with eigenvalue $k^2 > 0$ are \bea \psi_k(\rho) &
= & \frac{A(k)}{\sqrt{\pi}} \sin (\sqrt{k^2+U_0}~\rho) ~~~~~~~~~~0
\leq \rho \leq 1 \nn \\ \psi_k(\rho) & = &
\frac{1}{\sqrt{\pi}}\sin(k\rho + \theta(\rho)) ~~~~~~~~~~\leq \rho
\leq \infty
\label{12-1}
\eea
where the constants $A(k)$ and $\theta(k)$ are determined by matching at $\rho = 1$,
\bea
A(k) &= & \frac{k}{\sqrt{k^2
    \sin^2(\sqrt{k^2+U_0})+\sqrt{k^2+U_0}\cos^2(\sqrt{k^2+U_0}})}\nn \\ \theta(k)&=&
\tan ^{-1}\left(\frac{k \tan
  \left(\sqrt{k^2+U_0}\right)}{\sqrt{k^2+U_0}}\right)-k.
\label{12-2}
\eea
The solution for $k=0$ is
\bea
\psi_0(\rho) & = & \frac{B}{\sqrt{\pi}} \sin (\sqrt{U_0}~\rho) ~~~~~~~~~~0 \leq \rho \leq 1 \nn \\  
\psi_0(\rho) & = & a\rho + b ~~~~~~~~~~~~~~\leq \rho \leq \infty
\label{12-3}
\eea
The matching conditions at $\rho = 1$ now yield
\bea
\frac{B}{\sqrt{\pi}} \sin (\sqrt{U_0}~\rho) & = & a+b \nn \\
\frac{B\sqrt{U_0}}{\sqrt{\pi}} \cos (\sqrt{U_0}~\rho) & = & a
\label{12-4}
\eea
For any $a \neq 0$ the solution blows up at $\rho = \infty$. Thus regular solutions require $a = 0$. However the second equation in (\ref{12-4}) then imply that
\ben
\sqrt{U_0} = (n +\frac{1}{2}) \pi
\label{12-5}
\een
These are the zero modes. As we increase the depth of the potential, the first zero mode appears at $\sqrt{U_0} = \pi/2$. In the context of our model this is the potential where we have a critical point.

The small $k$ behavior of $A(k)$ and $\theta (k)$ can be read off from the expressions (\ref{12-2}). For a generic $U_0$ these are
\bea
A(k) & \sim & \frac{k}{\sqrt{U_0} \cos \sqrt{U_0}} + O(k^2) \nn \\
\theta (k) & \sim & k [ \frac{\tan \sqrt{U_0}}{\sqrt{U_0}} - 1] +O(k^3)
\label{12-6}
\eea
whereas for critical potentials we have
\bea
A(k) & \sim & 1 - \frac{k^2}{8} + O(k^4) \nn \\
\theta (k) & \sim & -\frac{\pi}{2} - \frac{k}{2}
\label{12-7}
\eea
Thus the small-$k$ behavior of the eigenfunctions are drastically different for the critical potentials. This has important implications for the coefficients like
$({\cal J}_{k}-{\cal J}_{0}),  (b_{k0}-b_{00} )$ and $(C_{k000}-C_{0000})$ in (\ref{scalingeqn2}). Consider for example the quantity ${\cal J}_k$. This is an integral of the form 
\ben
\int_0^\infty d\rho~J(\rho)~\chi_k (\rho)
\een
where $J(\rho)$ is a smooth function (which is $[-\partial_\rho^2+
V_0(\rho)] \ch_l $). If we replace the true eigenfunctions by those of our simplified problem, we get
\ben
{\cal J}_k = A(k) \int_0^1 d\rho \sin (\sqrt{k^2+U_0}\rho) J(\rho) + \int_1^\infty d\rho\sin(k\rho + \theta(k)) 
\een
Using (\ref{12-6}) and (\ref{12-7}) we therefore see that
\ben
{\cal J}_k -{\cal J}_0 \sim k~~~~~~~k \rightarrow 0
\een
for generic potentials, whereas
\ben
{\cal J}_k -{\cal J}_0 \sim k^2~~~~~~~k \rightarrow 0
\een
for critical potentials. It is straightforward to see that the behavior of the other coefficients $(b_{k0}-b_{00} )$ and $(C_{k000}-C_{0000})$ are similar.

The small-$k$ behavior of the eigenfunctions for the potential which
is relevant for us. $V_0^c(\rho)$
is quite similar. This has been discussed in the Appendix C of
\cite{bdas}. Going back to (\ref{scalingeqn2}) we therefore see that
the small $v$ expansion is generically {\em not valid} since the
corrections diverge at small $k$. However for the critical potential,
$\tilde\eta_k$ remain finite as $k \rightarrow 0$ and the expansion in
powers of $v^{2/5}$ makes sense.

\subsection{Scaling relation}
\vspace{0.5cm}
The validity of this expansion means that
the solution for $\chi_s$ is of the form
\ben
 \chi_s(\rho,u) \approx v^{\frac{1}{5}}\tilde \xi_0 (v^{\frac{2}{5}}u) 
\chi_0(\rho)+v^{\frac{3}{5}} \int \tilde \eta_k(v^{\frac{2}{5}}u)
\chi_k(\rho) dk + \cdots
\een
Thus the dynamics for small $v$ is dominated by the zero mode in the
critical region. The equation for  $\tilde
\xi_0$ is, however, exactly the same as the equation for the order
parameter $\vp$ in the LG model in the previous subsection, after
going back to the original variables prior to the rescaling in (\ref{rescale}).
Thus in this region the system behaves as one with dynamical critical
exponent $z=2$.
Therefore
we conclude that to leading order in small $v$, we have
\ben
<\cO>(u,v)  \sim {\rm Lim}_{\rho \rightarrow 0} [\rho^{1-\Delta_+}
  \chi_s (\rho,u)] \sim v^{1/5} <\cO>(tv^{2/5},1)
\een
and the time scale behaves as $v^{-2/5}$. 

The fact that the dynamics in the critical region is governed by an
equation with a first order time derivative is made manifest in our
treatment using Eddington-Finkelstein coordinates. The whole analysis
can be of course performed in principle in the $(t,r)$
coordinates. However, we suspect that in this case one has to exercise
extreme care, just as one had to extract out a leading horizon
behavior in the linearized problem of fields in a black brane
background \cite{starinets}.

Note that beyond the critical region, $J(u)$ departs from the form
$J(u) \sim vu$, and the expansion in powers of
$v^{2/5}$ is no longer valid. Now all the modes are important, and
integrating out higher modes can give rise to higher time derivatives
in the effective equation for $<\cO>$.

\subsection{Mass quench}
\vspace{0.5cm}
An analysis similar to the above can be carried out when the quenching
is performed by making the mass parameter of the bulk theory a
function of the retarded time $u$, keeping $J(u) = 0$. This does not have a direct
interpretation in the boundary field theory. However, as explained in
\cite{liu1}, the field $\phi$ can acquire  a mass because of a
coupling to some other field $\phi^\prime$. A time dependent boundary
value of the field $\phi^\prime$ can then lead to a time dependent
mass. When $m^2(u) \sim m_c^2 + vu$ near the critical point, we now
get a behavior $<\cO> \sim v^{1/4}$.  

\subsection{Numerical Results}
\begin{figure}[h]
\begin{center}
\includegraphics[scale=0.75]{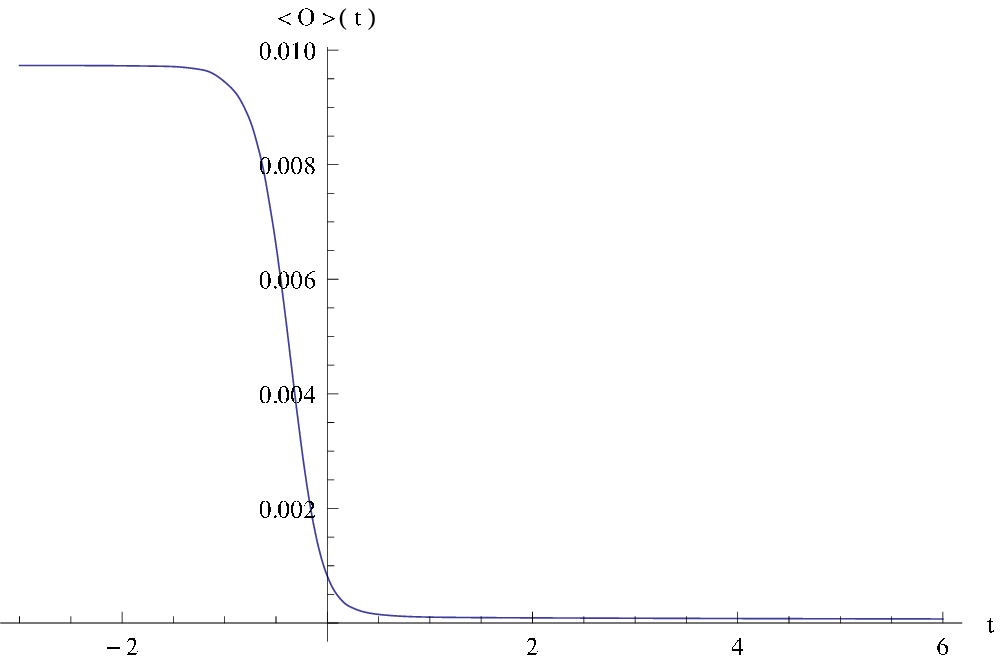}
\end{center}
 \caption{\label{fig:ord2}The order parameter $<\cO>(t)$ in boundary
theory.}
\end{figure}

\vspace{0.5cm}
We have performed some preliminary numerical work for the case of a
mass quench. Our results clearly display the propagation of
disturbances towards the horizon along a light cone
, and a decay of
the order parameter in a fashion similar to the $z=2$ LG
dynamics, which is shown in Figure(\ref{fig:ord2}). 
However our results are not accurate enough to verify the
scaling behavior found above. The late time decay should be governed
by the quasinormal modes \cite{hubeny2,quasi}.

\section{Outlook}
\vspace{0.5cm}
We have shown that holographic methods are useful in providing insight
into various questions related to quantum quench in strongly coupled
field theories which have a gravity dual. Interestingly, this is
possible in the probe approximation, which is typically much easier
to study. 

So far we have been able to study in some detail quench
dynamics near holographic critical points which are described by mean
field exponents. Not surprisingly, we found scaling behavior
characteristic of Landau-Ginsburg models with $z=2$.  This value of
the dynamical critical exponent is consistent with the results of 
\cite{holosupdynamical}.

It is important to perform our quench analysis for zero temperature,
where the equilibrium transition is of the BKT type \cite{liu1},
similar to brane models of zero temperature chiral symmetry breaking
transition in \cite{bktpapers}. This case is rather subtle, but can be
studied using similar methods.  We expect that in this case we will
have a $z=1$ theory dominating the critical region.
This would provide results for quench
dynamics which are not easily obtainable by other methods. Finally, a
more extensive numerical investigation should throw light on the
question of thermalization at late times, after the system has crossed
the critical region.

\section{Acknowledgements}
\vspace{0.5cm}
I would like to thank my collaborators Pallab Basu, Tatsuma Nishioka
and Tadashi Takayanagi for very enjoyable collaborations and many
insightful discussions. I would also like to thank Karl Landsteiner,
Satya Majumdar, Gautam Mandal, Shiraz Minwalla, Takeshi Morita,
Ganpathy Murthy, Omid Saremi, Alfred Shapere, Sandip Trivedi and
especially Kristan Jensen and Krishnendu Sengupta for
disucssions. S.R.D. would like to thank Institut de Fisica Teorica at
Madrid, Tata Institute of Fundamental Research at Mumbai and Indian
Association for the Cultivation of Science at Kolkata for hospitality
during the final stages of this work. Finally I thank the the organizers of 11th Workshop on Non-perturbative QCD in Paris, Sixth Crete Regional Meeting on String Theory in Milos and Quantum Theory and Symmetries 7 in Prague for organizing stimulating conferences.
This work is partially
supported by National Science Foundation grants PHY-0970069 and
PHY-0855614.

\end{document}